\documentclass[prx,twocolumn,showpacs]{revtex4}

\usepackage{graphicx}
\usepackage{dcolumn}
\usepackage{bm}
\usepackage{multirow}
\usepackage{amsmath,amssymb}

\begin{document}


\title{Emergence and spectral-weight transfer of electronic states in the Hubbard ladder}

\author{Masanori Kohno}
\email{KOHNO.Masanori@nims.go.jp}
\affiliation{National Institute for Materials Science, Tsukuba 305-0003, Japan}

\date{\today}

\begin{abstract}
The number of electronic bands is usually considered invariant regardless of the electron density in a band picture. 
However, in interacting systems, the spectral-weight distribution generally changes depending on the electron density, 
and electronic states can even emerge or disappear as the electron density changes. 
Here, to clarify how electronic states emerge and become dominant as the electron density changes, 
the spectral function of the Hubbard ladder with strong repulsion and strong intrarung hopping is studied 
using the non-Abelian dynamical density-matrix renormalization-group method. 
A mode emerging in the low-electron-density limit gains spectral weight as the electron density increases and governs the dimer Mott physics at quarter-filling. 
In contrast, the antibonding band, which is dominant in the low-electron-density regime, loses spectral weight and 
disappears at the Mott transition at half-filling, exhibiting the momentum-shifted magnetic dispersion relation in the small-doping limit. 
This paper identifies the origin of the electronic states responsible for the Mott transition and brings a new perspective to electronic bands 
by revealing the overall nature of electronic states over a wide energy and electron-density regime. 
\end{abstract}


\maketitle
\section{Introduction} 
In band theory, an electron is assumed to hop from one atomic orbital to another in an effective periodic potential, forming a band \cite{AshcroftMermin}; 
the number of bands is considered essentially determined by the number of atomic orbitals in a unit cell, 
which does not change with the electron density. 
In Fermi-liquid theory, electronic excitations other than the quasiparticle band are regarded 
as incoherent \cite{Nozieres}; the incoherent excitations are usually considered almost featureless and unimportant 
regardless of the electron density. 
\par
However, in interacting systems, some electronic excitations generally become dominant among many excited states, and 
the number of dominant modes can change depending on the electron density. 
Electronic excitations away from the Fermi level can also become dominant and exhibit significant characteristics. 
Thus, revealing the overall nature of electronic states over a wide energy and electron-density regime is important 
in the deeper understanding of the effects of strong electronic correlations. In particular, strong correlations significantly affect the electronic states near the Mott transition, 
which have attracted considerable attention in relation to high-temperature superconductivity \cite{DagottoRMP,ImadaRMP,KohnoRev}. 
\par
In this paper, to clarify how electronic states emerge, change, and disappear as the electron density changes in strongly correlated systems, 
the spectral function of the Hubbard ladder, which is one of the simplest models containing the essence of electronic correlations, 
is investigated in the regime of strong Coulomb repulsion and strong intrarung hopping. 
The qualitative features of the results would be generally true for coupled dimer systems, such as the dimer Mott insulators of molecular solids \cite{KinoEffU,SeoRev} 
regardless of the lattice structure or dimensionality 
as long as the Coulomb repulsion and intradimer hopping are much stronger than the interdimer hopping. 
In particular, the perturbative arguments shown in this paper can be straightforwardly extended to bilayer systems. 
\par
The main features we focus on in this paper are the (1) emergent electronic states in the low-electron-density regime [Sec. \ref{sec:lowDensity}], 
(2) spectral-weight transfer from the dominant modes to the emergent modes, 
which makes the emergent modes dominant, whereas the dominant modes significantly lose spectral weight as the electron density increases to half-filling [Sec. \ref{sec:overall}], 
(3) dimer Mott gap at quarter-filling, whose value is significantly limited by the intrarung hopping in the strong-Coulomb-repulsion regime [Sec. \ref{sec:quarterFilling}], 
and (4) emergent electronic states upon doping a Mott insulator by which the Mott transition is characterized [Sec. \ref{sec:MottTrans}]. 
The above features are contrasted with conventional views, such as a band picture. 
\section{Model and method} 
We consider the Hubbard ladder defined by the following Hamiltonian:
\begin{eqnarray}
\label{eq:Hamiltonian}
{\cal H}&=&-t_{\parallel}\sum_{i,\alpha,\sigma}
(c^{\alpha\dagger}_{i,\sigma}c^{\alpha}_{i+1,\sigma}+{\mbox {H.c.}})
-t_{\perp}\sum_{i,\sigma}(c^{1\dagger}_{i,\sigma}c^2_{i,\sigma}+{\mbox {H.c.}})\nonumber\\
&&+U\sum_{i,\alpha}n^{\alpha}_{i,\uparrow}n^{\alpha}_{i,\downarrow}-\mu\sum_{i,\alpha,\sigma}n^{\alpha}_{i,\sigma},
\end{eqnarray}
where $c^{\alpha}_{i,\sigma}$ and $n^{\alpha}_{i,\sigma}$, respectively, denote the annihilation and number operators of an electron with spin $\sigma(=\uparrow, \downarrow)$ at the site of leg $\alpha(=1, 2)$ and rung $i$. 
Hereafter, the number of sites in a leg, total number of sites, and number of electrons are denoted by $L$, $N_{\rm s}(=2L)$, and $N_{\rm e}$, respectively. 
The doping concentration $\delta$ is defined as $\delta=1-n$, where $n$ denotes the electron density $(n=N_{\rm e}/N_{\rm s})$. 
\par
The single-particle spectral function $A({\bm k},\omega)$ and 
the dynamical spin structure factor $S({\bm k},\omega)$ are defined as 
\begin{equation}
\begin{aligned}
A({\bm k},\omega)&=\left\{
\begin{array}{lll}
\frac{1}{2}\sum_{l,\sigma}|\langle l|c_{{\bm k},\sigma}^{\dagger}|{\rm GS}\rangle|^2\delta(\omega-\varepsilon_l)&\mbox{for}&\omega>0,\\
\frac{1}{2}\sum_{l,\sigma}|\langle l|c_{{\bm k},\sigma}|{\rm GS}\rangle|^2\delta(\omega+\varepsilon_l)&\mbox{for}&\omega<0,
\end{array}\right.\\
S({\bm k},\omega)&=\frac{1}{3}\sum_{l,\gamma}|\langle l|S_{\bm k}^\gamma|{\rm GS}\rangle|^2\delta(\omega-\varepsilon_l),
\end{aligned}
\end{equation}
where $\varepsilon_l$ represents the excitation energy of the eigenstate $|l\rangle$ from the ground state $|{\rm GS}\rangle$. 
Here, $c^{\dagger}_{{\bm k},\sigma}$ and $S_{\bm k}^\gamma$ denote the Fourier transform of $c^{\dagger}_{i,\sigma}$ and 
that of the $\gamma(=x,y,z)$ component of the spin operator ${\bm S}_i$, respectively. 
For ladders, ${\bm k}=(k_{\parallel},k_{\perp})$, 
where $k_{\parallel}$ and $k_{\perp}$ denote the momenta in the leg and rung directions, respectively. 
Because $\{c_{{\bm k},\sigma},c_{{\bm k},\sigma}^{\dagger}\}=1$, 
$A({\bm k},\omega)$ satisfies the following sum rule at each ${\bm k}$: 
\begin{equation}
\label{eq:sumrule}
\int_{-\infty}^\infty d\omega A({\bm k},\omega)=1.
\end{equation}
\par
We consider the case of $0\le n\le 1$ without loss of generality 
because $A({\bm k},\omega)$ for $1<n\le 2$ can be obtained as $A({\bm k}+{\bm \pi},-\omega)$ at the electron density $2-n$ by using the particle-hole transformation. 
The Hubbard ladder has been studied primarily on the ground-state properties \cite{Balents,NoackPRL,NoackPairing,TroyerPairing,singleHole}, 
spin and charge excitations \cite{Endres,SkwNkwDMRGFLEX,spinGap}, spectral function around half-filling \cite{NoackPairing,Endres,Feiguin}, 
charge and photo dynamics \cite{chargeDynamics,photoDynamics}, and ferromagnetism \cite{KohnoUinf,KrivnovUinf,Kivelson}. 
In this paper, to clarify the evolution of electronic states as a function of the electron density, 
we investigate the spectral function in the overall electron-density regime primarily for $U\gg t_{\perp}\gg t_{\parallel}>0$ 
($U t_{\parallel}/t_{\perp}^2$ is not too large for the ground state to have spin 0 or 1/2 \cite{KohnoUinf,KrivnovUinf}) 
based on the numerical results for $U/t_{\parallel}=16$ and $t_{\perp}/t_{\parallel}=2$ obtained 
using the non-Abelian dynamical density-matrix renormalization-group (DDMRG) method \cite{DDMRG,nonAbelianHub,nonAbeliantJ,nonAbelianThesis,KohnoRev,Kohno2DtJ,KohnoDIS}. 
The DDMRG calculations were performed on a 120-site cluster under open boundary conditions with 240 states retained for the density matrix. 
The truncation errors are negligibly small in the scales used for the figures in this paper. 
\section{Overall spectral features} 
\label{sec:overall}
\begin{figure*}
\includegraphics[width=17cm]{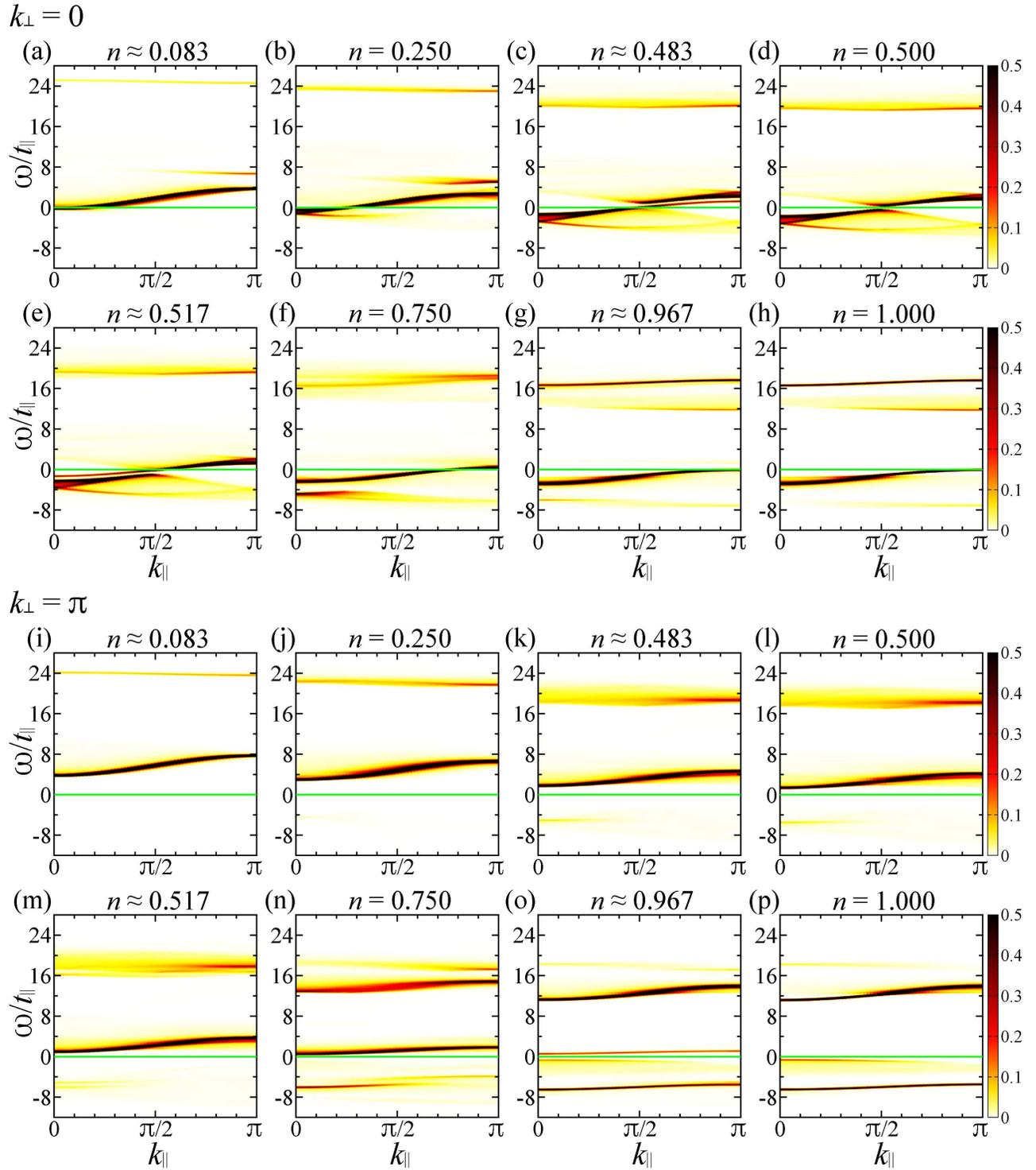}
\caption{$A({\bm k},\omega)t_{\parallel}$ for [(a)--(h)] $k_{\perp}=0$ and [(i)--(p)] $\pi$ 
at [(a), (i)] $n\approx 0.083$, [(b), (j)] 0.25, [(c), (k)] 0.483, [(d), (l)] 0.5, [(e), (m)] 
0.517, [(f), (n)] 0.75, [(g), (o)] 0.967, and [(h), (p)] 1 for $U/t_{\parallel}=16$ and $t_{\perp}/t_{\parallel}=2$ obtained using the non-Abelian DDMRG method. 
The green lines indicate $\omega=0$. Gaussian broadening with a standard deviation of $0.1t_{\parallel}$ has been used.} 
\label{fig:U16}
\end{figure*}
The spectral-weight distributions of electronic states from the low-electron-density regime ($n\approx 0.083$) to half-filling ($n=1$) are 
shown in Fig. \ref{fig:U16} for $k_{\perp}=0$ [Figs. \ref{fig:U16}(a)--\ref{fig:U16}(h)] and $k_{\perp}=\pi$ [Figs. \ref{fig:U16}(i)--\ref{fig:U16}(p)]. 
One might naively expect that the dominant modes in the low-electron-density regime 
[$0\lesssim\omega/t_{\parallel}\lesssim 4$ in Fig. \ref{fig:U16}(a); $4\lesssim\omega/t_{\parallel}\lesssim 8$ in Fig. \ref{fig:U16}(i)] 
continuously deform into those at half-filling 
[$-3\lesssim\omega/t_{\parallel}\lesssim 0$ in Fig. \ref{fig:U16}(h); $11\lesssim\omega/t_{\parallel}\lesssim 14$ in Fig. \ref{fig:U16}(p)] 
as the electron density increases. However, they are different in origin. 
\par
\begin{figure}
\includegraphics[width=8.3cm]{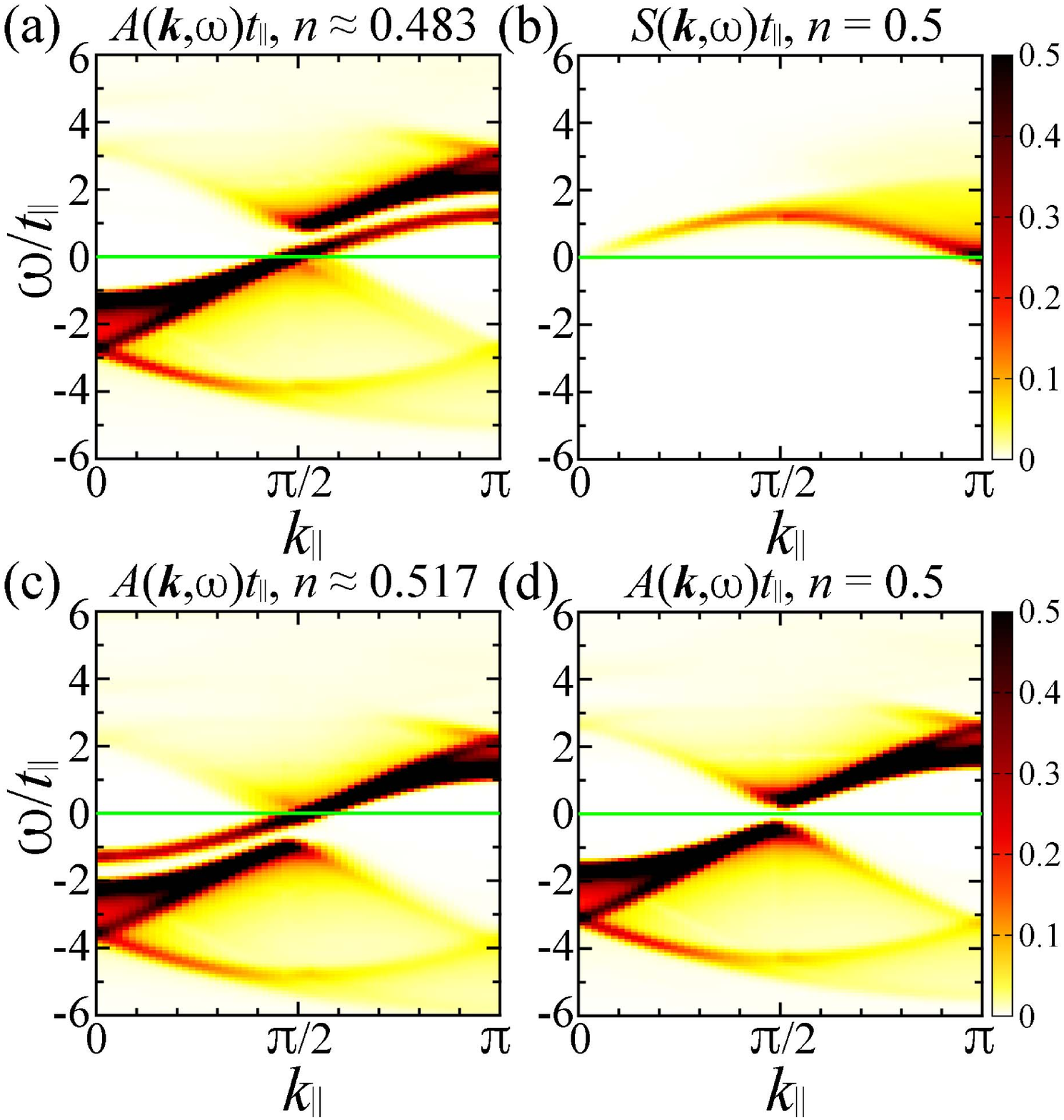}
\caption{(a), (c), (d) $A({\bm k},\omega)t_{\parallel}$ for $k_{\perp}=0$ at [(a)] $n\approx 0.483$, [(c)] $n\approx 0.517$, and [(d)] $n= 0.5$ 
for $U/t_{\parallel}=16$ and $t_{\perp}/t_{\parallel}=2$ obtained using the non-Abelian DDMRG method 
[closeup of Figs. \ref{fig:U16}(c), \ref{fig:U16}(e), and \ref{fig:U16}(d) near the Fermi level, respectively]. 
(b) $S({\bm k},\omega)t_{\parallel}$ for $k_{\perp}=0$ at $n=0.5$ 
for $U/t_{\parallel}=16$ and $t_{\perp}/t_{\parallel}=2$ obtained using the non-Abelian DDMRG method. 
The green lines indicate $\omega=0$. Gaussian broadening with a standard deviation of $0.1t_{\parallel}$ has been used.} 
\label{fig:quarterFilling}
\end{figure}

The dominant mode in the low-electron-density regime for $k_{\perp}=0$ [Fig. \ref{fig:U16}(a)] gradually loses spectral weight 
with the electron density [Figs. \ref{fig:U16}(b) and \ref{fig:U16}(c)]. 
At quarter-filling ($n=1/2$), this mode is located below $\omega=0$, separated by a small gap from the mode above $\omega=0$ [Figs. \ref{fig:U16}(d) and \ref{fig:quarterFilling}(d)]. 
The spectral weight further decreases and almost disappears at half-filling [Figs. \ref{fig:U16}(e)--\ref{fig:U16}(h)]. 
\par
As for $k_{\perp}=\pi$, the dominant mode in the low-electron-density regime [Fig. \ref{fig:U16}(i)] gradually loses spectral weight 
with the electron density [Figs. \ref{fig:U16}(j)--\ref{fig:U16}(o)], and completely disappears at half-filling [Fig. \ref{fig:U16}(p)]. 
\par
Instead, because of the sum rule [Eq. (\ref{eq:sumrule})], emergent modes in the low-electron-density regime [Figs. \ref{fig:U16}(a) and \ref{fig:U16}(i)] 
gradually gain spectral weight [Figs. \ref{fig:U16}(b)--\ref{fig:U16}(g), \ref{fig:U16}(j)--\ref{fig:U16}(o)], 
and become dominant at half-filling [Figs. \ref{fig:U16}(h) and \ref{fig:U16}(p)]. 
Thus, a significant amount of the spectral weight transfers from the dominant modes to the emergent modes as the electron density increases. 
\par
It should be noted that not only the high-energy modes of $O(U)$ but also an intermediate-energy mode emerges 
[$\omega/t_{\parallel}\approx 7$ and $k_{\parallel}\approx\pi$ in Fig. \ref{fig:U16}(a)], 
which is separated from the low-energy bonding band by an energy gap, 
and becomes the most dominant at half-filling for $k_{\perp}=0$ [$-3\lesssim\omega/t_{\parallel}\lesssim 0$ in Fig. \ref{fig:U16}(h)]. 
\par
These features are due to strong electronic correlations and contrast with a rigid-band picture 
in which the bonding and antibonding bands remain dominant regardless of the electron density. 
\section{Zero electron density} 
\begin{figure}
\includegraphics[width=8.3cm]{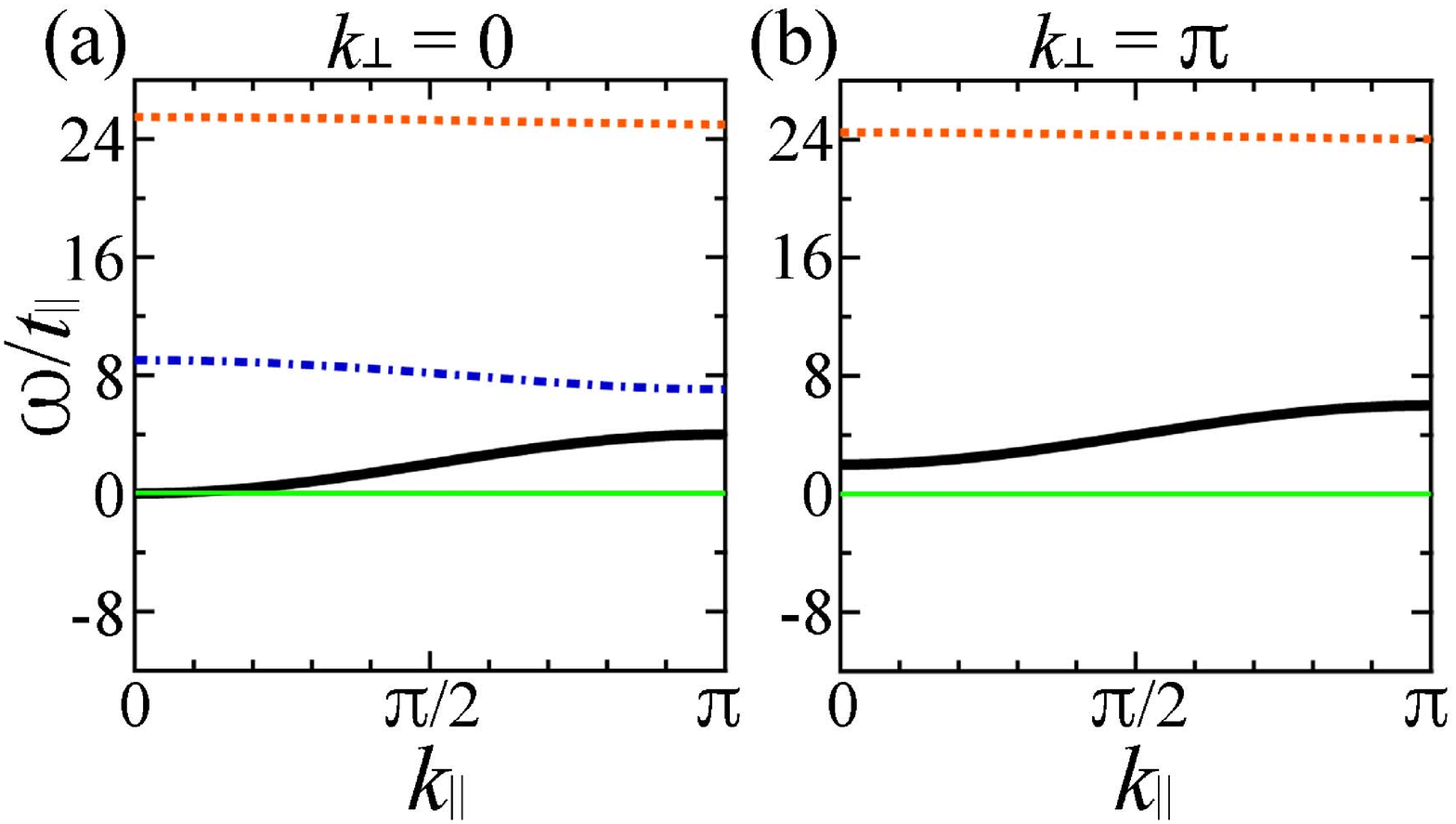}
\caption{Dispersion relation of electronic excitations in the low-electron-density limit for $U/t_{\parallel}=16$ and $t_{\perp}/t_{\parallel}=2$ at $k_{\perp}=0$ [(a)] and $\pi$ [(b)]. 
The dotted orange curves indicate the high-energy solutions of Eq. (\ref{eq:2ele}). 
The solid black curves indicate $\epsilon^0_{\bm k}$ [Eq. (\ref{eq:e0k})]. 
The dashed-dotted blue curve in (a) indicates the high-energy solutions of Eq. (\ref{eq:2ele}) 
for the effective Hubbard chain with $U^-_{\rm eff}=E_-+2t_{\perp}$ [Eq. (\ref{eq:effU})]. 
The green lines indicate $\omega=0$.} 
\label{fig:2e}
\end{figure}

To understand the nature of such complicated spectral features, we consider the properties of characteristic modes from the low-electron-density side. 
At zero electron density ($n=0$), the electron-addition spectra show the noninteracting dispersion relations 
because the interaction term does not work with the added electron. 
Thus, the Hamiltonian ($U=0$) can be diagonalized in the momentum space as 
\begin{equation}
{\cal H}_0=\sum_{{\bm k},\sigma}\epsilon^0_{\bm k}
c^{\dagger}_{{\bm k},\sigma}c_{{\bm k},\sigma}. 
\end{equation}
The noninteracting dispersion relation $\epsilon^0_{\bm k}$ is obtained as 
\begin{equation}
\label{eq:e0k}
\epsilon^0_{\bm k}=-2t_{\parallel}\cos k_{\parallel}-t_{\perp}\cos k_{\perp}-\mu, 
\end{equation}
where $k_{\perp}=0$ for the bonding band and $k_{\perp}=\pi$ for the antibonding band (solid black curves in Fig. \ref{fig:2e}). 
In the low-electron-density limit ($n\rightarrow 0$), $\mu\rightarrow -2t_{\parallel}-t_{\perp}$. 
\section{Low electron density} 
\label{sec:lowDensity}
\subsection{High-energy emergent modes} 
\label{sec:highEmodes}
In addition to the dominant modes originating from the noninteracting bonding and antibonding bands [Eq. (\ref{eq:e0k})], 
small spectral weights emerge when the electron density becomes nonzero [Figs. \ref{fig:U16}(a) and \ref{fig:U16}(i)]. 
The emergent modes at high energies of $O(U)$ can be regarded as the upper Hubbard bands. 
To capture their characteristics, we consider a two-electron system. 
The energy of a spin-singlet eigenstate $\epsilon_{\bm k}$ is generally obtained 
as a solution of the following equation \cite{Hubbard2ele}: 
\begin{equation}
\label{eq:2ele}
1=\frac{U}{N_{\rm s}}\sum_{\bm p}\frac{1}{\epsilon_{\bm k}-\epsilon^0_{{\bm k}-{\bm p}}-\epsilon^0_{\bm p}}. 
\end{equation}
By expanding Eq. (\ref{eq:2ele}) in powers of $U$, 
the dispersion relations of the high-energy modes for $U\gg t_{\perp}$, $t_{\parallel}$ can be obtained as 
\begin{equation}
\label{eq:2eleLadder}
\begin{aligned}
\epsilon_{(k_{\parallel},0)}&=U+2t_{\perp}+4t_{\parallel}
+J_{\parallel}\left(\cos k_{\parallel}+1\right)+J_{\perp},\\
\epsilon_{(k_{\parallel},\pi)}&=U+2t_{\perp}+4t_{\parallel}
+J_{\parallel}\left(\cos k_{\parallel}+1\right),
\end{aligned}
\end{equation}
up to $O(t_{\parallel}^2/U)$ and $O(t_{\perp}^2/U)$, where $J_{\parallel}=4t^2_{\parallel}/U$ and $J_{\perp}=4t^2_{\perp}/U$ 
(dotted orange curves in Fig. \ref{fig:2e}). 
\par
The above results can also be explained in terms of the modes of double occupancy. 
We define $|\psi^+(k_{\parallel})\rangle$ and $|D^-(k_{\parallel})\rangle$ for $k_{\perp}=0$ and $\pi$, respectively, as 
\begin{equation}
|X(k_{\parallel})\rangle=\frac{1}{\sqrt{L}}\sum_{j=1}^L\mathrm{e}^{i k_{\parallel}r_j}
|X\rangle_j\prod_{l\ne j}^{L}|0\rangle_l, 
\label{eq:mode0}
\end{equation}
where $r_j$ denotes the coordinate of rung $j$ in the leg direction, and $X$ represents $\psi^+$ and $D^-$. 
Here, $|\psi^+\rangle_j$, $|D^-\rangle_j$, and $|0\rangle_j$ denote the eigenstates of the $j$th rung defined in Table \ref{tbl:rung}. 
\begin{table}
\caption{Eigenstates and energies on a rung.}
\begin{tabular}{ll}
\hline\hline
Eigenstate\footnotemark[0]&Energy\\\hline
$|0\rangle=|0,0\rangle$&$E_0=0$\\
$|B_{\sigma}\rangle=(|\sigma,0\rangle+|0,\sigma\rangle)/\sqrt{2}$&$E_B=-t_{\perp}-\mu$\\
$|A_{\sigma}\rangle=(|\sigma,0\rangle-|0,\sigma\rangle)/\sqrt{2}$&$E_A=t_{\perp}-\mu$\\
$|\psi^+\rangle=-\zeta_-|S\rangle+\zeta_+|D^+\rangle$&$E_{\psi^+}=E_+-2\mu$\\
$|\psi^-\rangle=\zeta_+|S\rangle+\zeta_-|D^+\rangle$&$E_{\psi^-}=E_--2\mu$\\
$|T^+\rangle=|\uparrow,\uparrow\rangle$&$E_{T^+}=-2\mu$\\
$|T^-\rangle=|\downarrow,\downarrow\rangle$&$E_{T^-}=-2\mu$\\
$|T^0\rangle=(|\uparrow,\downarrow\rangle+|\downarrow,\uparrow\rangle)/\sqrt{2}$&$E_{T^0}=-2\mu$\\
$|D^-\rangle$&$E_{D^-}=U-2\mu$\\
$|G_{\sigma}\rangle=(|\sigma,\uparrow\downarrow\rangle+|\uparrow\downarrow,\sigma\rangle)/\sqrt{2}$&
$E_G=U+t_{\perp}-3\mu$\\
$|F_{\sigma}\rangle=(|\sigma,\uparrow\downarrow\rangle-|\uparrow\downarrow,\sigma\rangle)/\sqrt{2}$&
$E_F=U-t_{\perp}-3\mu$\\
$|W\rangle=|\uparrow\downarrow,\uparrow\downarrow\rangle$&$E_W=2U-4\mu$\\\hline\hline
\end{tabular}
\footnotetext[0]{$|S\rangle=(|\uparrow,\downarrow\rangle-|\downarrow,\uparrow\rangle)/\sqrt{2}$, 
$|D^{\pm}\rangle=(|\uparrow\downarrow,0\rangle\pm|0,\downarrow\uparrow\rangle)/\sqrt{2}$,\\
\quad$\zeta_{\pm}=\sqrt{\frac{1}{2}\left(1\pm U/\sqrt{U^2+16t_{\perp}^2}\right)}$, 
$E_{\pm}=\frac{U}{2}\pm\frac{\sqrt{U^2+16t_\perp^2}}{2}$.}
\label{tbl:rung}
\end{table}
The effective hoppings of $|\psi^+\rangle$ and $|D^-\rangle$ are obtained 
using second-order perturbation theory with respect to $t_{\parallel}$ as 
\begin{equation}
\begin{aligned}
t_{\parallel{\rm eff}}^{\psi^+}&=-\frac{4t_{\parallel}^2}{U}+\frac{2t_{\parallel}^2}{\sqrt{U^2+16t_{\perp}^2}},\\
t_{\parallel{\rm eff}}^{D^-}&=-\frac{2t_{\parallel}^2}{U},
\end{aligned}
\end{equation}
respectively, which reduce to 
\begin{equation}
t_{\parallel{\rm eff}}^{\psi^+}\approx t_{\parallel{\rm eff}}^{D^-}\approx-\frac{J_{\parallel}}{2},
\end{equation}
for $U\gg t_{\perp}$. 
By taking into account the bond energy between $|\psi^+\rangle$ and $|0\rangle$, $\xi_{\psi^+0}(=-t_{\parallel{\rm eff}}^{\psi^+})$, and 
that between $|D^-\rangle$ and $|0\rangle$, $\xi_{D^-0}(=-t_{\parallel{\rm eff}}^{D^-})$, 
the energies of the high-energy modes are obtained as
\begin{equation}
\label{eq:highEmodes}
\begin{aligned}
E_{\parallel{\rm eff}}^{\psi^+}&\approx J_{\parallel}(\cos k_{\parallel}+1)+E_+-2\mu,\\
E_{\parallel{\rm eff}}^{D^-}&\approx J_{\parallel}(\cos k_{\parallel}+1)+U-2\mu,
\end{aligned}
\end{equation}
up to $O(t_{\parallel}^2/U)$ for $U\gg t_{\perp}\gg t_{\parallel}$. 
Here, 
\begin{equation}
E_{\pm}=\frac{U}{2}\pm\frac{\sqrt{U^2+16t_\perp^2}}{2},
\end{equation}
which reduces to 
\begin{equation}
\label{eq:effEpm}
\begin{aligned}
E_+&\approx U+J_{\perp},\\
E_-&\approx -J_{\perp},
\end{aligned}
\end{equation}
for $U\gg t_{\perp}$. 
By putting $\mu=-2t_{\parallel}-t_{\perp}$ in the low-electron-density limit (the ground-state energy of the one-electron system is zero), 
Eq. (\ref{eq:highEmodes}) reduces to Eq. (\ref{eq:2eleLadder}) for $U\gg t_{\perp}\gg t_{\parallel}$. 
This result implies that the high-energy modes in the low-electron-density limit can be interpreted as 
the modes of double occupancy [Eq. (\ref{eq:mode0})] for $U\gg t_{\perp}\gg t_{\parallel}$. 
\subsection{Intermediate-energy emergent mode} 
\label{sec:intermediateModes}
As mentioned in Sec. \ref{sec:overall}, not only the high-energy modes of $O(U)$ 
but also an intermediate-energy mode emerges [$\omega/t_{\parallel}\approx 7$ and $k_{\parallel}\approx\pi$ in Fig. \ref{fig:U16}(a)]. 
To clarify the nature of this mode, we consider an effective model for $U$, $t_{\perp}\gg t_{\parallel}$. 
In the low-electron-density regime, an electron with $k_{\perp}=0$ on a rung, $|B_{\sigma}\rangle$, hops almost freely along the leg. 
When two electrons with opposite spins sit on the same rung, they are excited to one of the two-electron eigenstates with $k_{\perp}=0$ on a rung, $|\psi^{\pm}\rangle$ 
(Table \ref{tbl:rung}). 
Thus, the effective model for $k_{\perp}=0$ can be obtained as the Hubbard chain 
with the following effective interaction $U^{\pm}_{\rm eff}$ \cite{KinoEffU} and effective chemical potential $\mu_{\rm eff}$: 
\begin{equation}
\label{eq:effU}
\begin{aligned}
U^{\pm}_{\rm eff}&=E_{\pm}+2t_{\perp},\\
\mu_{\rm eff}&=\mu+t_{\perp},
\end{aligned}
\end{equation}
by equating $E_B$ and $E_{\psi^\pm}$ with the effective single-site energies $-\mu_{\rm eff}$ and $U^{\pm}_{\rm eff}-2\mu_{\rm eff}$, respectively. 
\par
In the Hubbard chain, the dispersion relation of the high-energy mode in a two-electron system can also be obtained from Eq. (\ref{eq:2ele}) as 
\begin{equation}
\label{eq:2eleChain}
\epsilon_{k_{\parallel}}=U+4t_{\parallel}+J_{\parallel}\left(\cos k_{\parallel}+1\right), 
\end{equation}
up to $O(t_{\parallel}^2/U$) in the large-$U/t_{\parallel}$ regime [$t_{\perp}=J_{\perp}=0$ in Eq. (\ref{eq:2eleLadder})]. 
By putting $U^{\pm}_{\rm eff}$ [Eqs. (\ref{eq:effEpm}) and (\ref{eq:effU})] into Eq. (\ref{eq:2eleChain}), we obtain 
\begin{equation}
\begin{aligned}
\epsilon^+_{(k_{\parallel},0)}&=U+&2t_{\perp}+J_{\perp}+4t_{\parallel}+J_{\parallel}\left(\cos k_{\parallel}+1\right),\\
\epsilon^-_{(k_{\parallel},0)}&=&2t_{\perp}-J_{\perp}+4t_{\parallel}+J_{\parallel}\left(\cos k_{\parallel}+1\right), 
\end{aligned}
\end{equation}
up to $O(t_{\perp}^2/U)$ and $O(t_{\parallel}^2/U)$ for $U\gg t_{\perp}\gg t_{\parallel}$. 
The former [$\epsilon^+_{(k_{\parallel},0)}$] corresponds to 
the high-energy mode of $O(U)$ [Eqs. (\ref{eq:2eleLadder}) and (\ref{eq:highEmodes})]. 
The latter [$\epsilon^-_{(k_{\parallel},0)}$] corresponds to the emergent mode in the intermediate-energy regime 
[$\omega/t_{\parallel}\approx 7$ and $k_{\parallel}\approx\pi$ in Fig. \ref{fig:U16}(a); dashed-dotted blue curve in Fig. \ref{fig:2e}(a)]. 
\par
\begin{figure}
\includegraphics[width=8.3cm]{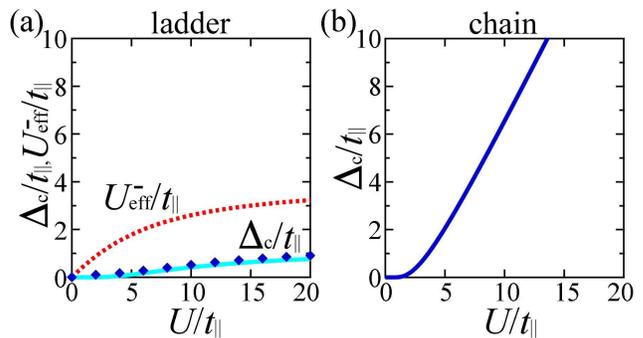}
\caption{(a) $U^-_{\rm eff}/t_{\parallel}$ for $t_{\perp}/t_{\parallel}=2$ [Eq. (\ref{eq:effU})] (dotted red curve). 
Charge gap $\Delta_{\rm c}$ of the Hubbard ladder at $n=1/2$ for $t_{\perp}/t_{\parallel}=2$ 
determined as the chemical-potential difference obtained using the non-Abelian DDMRG method in a 120-site cluster (blue diamonds), 
and $\Delta_{\rm c}$ of the effective Hubbard chain with $U^-_{\rm eff}$ at $n_{\rm eff}(=2n)=1$ (solid cyan curve) obtained using (b). 
(b) $\Delta_{\rm c}$ of the Hubbard chain at $n=1$ obtained using the Bethe ansatz \cite{LiebWu}.}
\label{fig:Gapn05}
\end{figure}

From the above analysis, the emergent mode in the intermediate-energy regime can be interpreted 
as the upper Hubbard band of the effective Hubbard chain with $U^-_{\rm eff}=E_-+2t_{\perp}\approx -J_{\perp}+2t_{\perp}$ 
[dotted red curve in Fig. \ref{fig:Gapn05}(a)]. 
Thus, the energy of this mode is not $O(U)$ but $O(t_{\perp})$ for $U\gg t_{\perp}\gg t_{\parallel}$. 
\subsection{Remarks on the upper Hubbard band} 
In a conventional band picture, the splitting of a band into upper and lower bands is considered a result of symmetry breaking. 
For example, antiferromagnetic ordering, which causes folding of the Brillouin zone, 
is considered responsible for the formation of the gap in an antiferromagnetic insulator in a band picture. 
Nevertheless, the emergence of the high-energy states is a general characteristic of strongly interacting systems on a lattice, 
which does not require symmetry breaking or long-range order. 
\par
The mechanism of the emergence of the upper Hubbard band can, instead, be interpreted as the formation of a pair or 
a bound state \cite{Hubbard2ele,Kohno1DHub,Kohno1DHeisH,KohnoUHB,Kohno1DSF,Pereira1DSFDMRG}. 
The simplest case is the limit of the low-electron-density and strong repulsion where the high-energy mode can be interpreted 
as a mode of double occupancy (a bound state of electrons with opposite spins \cite{Hubbard2ele}) [Eq. (\ref{eq:mode0})] as shown in Sec. \ref{sec:highEmodes}. 
More generally, the interpretation as the formation of a pair can be justified in terms of string solutions in one-dimensional (1D) systems. 
Among Bethe-ansatz solutions, there are solutions involving a string which can be regarded as a pair of particles \cite{TakahashiHub,BetheAnsatz}. 
In the Hubbard chain, the upper Hubbard band has been identified as the $k$-$\Lambda$ string solutions \cite{Kohno1DHub}. 
Similarly, the high-energy states of the antiferromagnetic Heisenberg chain in a magnetic field in $S^{+-}(k,\omega)$ 
(excitation of flipping a majority spin to a minority spin) have been identified as the two-string solutions \cite{Kohno1DHeisH}, 
which correspond to the upper Hubbard band of the interacting hard-core bosons \cite{KohnoUHB} and interacting spinless fermions on a chain \cite{Kohno1DSF,Pereira1DSFDMRG}. 
Thus, these high-energy states can be identified as states involving a pair of particles. 
\par
It should be noted that the quasiparticle responsible for the upper Hubbard band is not exactly the double occupancy, in general. 
In fact, the double occupancy exists even in the ground state at half-filling for $U<\infty$, 
and the double occupancy is not specified by a quantum number of eigenstates. 
In the Hubbard chain, the quasiparticle responsible for the upper Hubbard band has been identified 
in terms of the quantum number for the $k$-$\Lambda$ string \cite{Kohno1DHub} 
similarly to the spinon and holon (defined in terms of the quantum numbers for spin and charge, respectively). 
In higher dimensions, the quasiparticle responsible for the upper Hubbard band could be interpreted as a pair of particles \cite{Hubbard2ele} or 
that of a chain deformed by interchain hopping (Sec. \ref{sec:intermediateModes}) \cite{Kohno2DHub,KohnoQ1DHeisH}. 
\section{Quarter-filling} 
\label{sec:quarterFilling}
At quarter-filling ($n=1/2$), the system becomes a Mott insulator [Figs. \ref{fig:U16}(d), \ref{fig:U16}(l), and \ref{fig:quarterFilling}(d)] 
because the Mott transition occurs at $n_{\rm eff}(=N_{\rm e}/L=2n)=1$ in the effective Hubbard chain with $U^-_{\rm eff}$ for $k_{\perp}=0$ [Sec. \ref{sec:intermediateModes}]. 
\par
To understand the nature of this dimer Mott insulator, 
we consider the electronic states of the upper Hubbard band of the effective Hubbard chain. 
As mentioned in Sec. \ref{sec:intermediateModes}, the two values of the effective $U$, $U^{\pm}_{\rm eff}=E_{\pm}+2t_{\perp}$ [Eq. (\ref{eq:effU})], 
reflect the two doubly occupied states on a rung (Table \ref{tbl:rung}), 
\begin{equation}
\begin{aligned}
|\psi^+\rangle&=-&\zeta_-|S\rangle+\zeta_+|D^+\rangle,\\
|\psi^-\rangle&=&\zeta_+|S\rangle+\zeta_-|D^+\rangle,
\end{aligned}
\end{equation}
where 
\begin{equation}
\zeta_{\pm}=\sqrt{\frac{1}{2}\left(1\pm \frac{U}{\sqrt{U^2+16t_{\perp}^2}}\right)}. 
\end{equation}
In the large-$U/t_{\perp}$ limit, $|\psi^+\rangle\rightarrow |D^+\rangle$ and $|\psi^-\rangle\rightarrow |S\rangle$ 
because $\zeta_+\rightarrow 1$ and $\zeta_-\rightarrow 0$. 
Hence, the high-energy mode of $O(U)$ has a larger component of the doubly occupied sites 
[$|D^+\rangle=(|\uparrow\downarrow,0\rangle+|0,\downarrow\uparrow\rangle)/\sqrt{2}$], 
whereas the low-energy mode of $O(t_{\perp})$ has a larger component of the spin-singlet state without doubly occupied sites 
[$|S\rangle=(|\uparrow,\downarrow\rangle-|\downarrow,\uparrow\rangle)/\sqrt{2}$] in the large-$U/t_{\perp}$ regime. 
Thus, in the dimer Mott insulator whose gap is essentially determined by $U_{\rm eff}^-$, 
the doubly occupied state $|\psi^-\rangle$ can primarily be regarded as the spin-singlet state without doubly occupied sites on a rung; 
the Mott gap is not of $O(U)$ but of $O(t_{\perp})$ ($U_{\rm eff}^-=E_-+2t_{\perp}\approx -J_{\perp}+2t_{\perp}$) for $U\gg t_{\perp}\gg t_{\parallel}$ 
[blue diamonds and solid cyan curve in Fig. \ref{fig:Gapn05}(a)] \cite{KohnoUinf}. 
\par
The Mott gap and the effective $U$ are relevant not only to the charge excitation but also to spin fluctuations and electronic excitations. 
The effective spin coupling at quarter-filling is $J^-_{\parallel{\rm eff}}=4t_{\parallel}^2/U^-_{\rm eff}$ $(\gg J_{\parallel}=4t_{\parallel}^2/U)$ for $U\gg t_{\perp}\gg t_{\parallel}$. 
Hence, the spin degrees of freedom for $k_{\perp}=0$ can be described as the effective Heisenberg chain with $J^-_{\parallel{\rm eff}}$, 
which shows a two-spinon continuum: $\omega=\frac{\pi J^-_{\parallel{\rm eff}}}{2}(\sin p_{\parallel}^1+\sin p_{\parallel}^2)$, 
where $k_{\parallel}=p_{\parallel}^1+p_{\parallel}^2$ for $0\le p_{\parallel}^1<p_{\parallel}^2\le \pi$, 
and the spin-wave mode at the lower edge: $\omega=\frac{\pi J^-_{\parallel{\rm eff}}}{2}|\sin k_{\parallel}|$ \cite{desCloizeaux} [Fig. \ref{fig:quarterFilling}(b)]. 
\par
By doping the dimer Mott insulator with a hole (an electron), the spin-wave mode emerges in the electronic-excitation spectrum 
with the dispersion relation shifted by the Fermi momenta $k_{\parallel {\rm F}}=\pm\pi/2$ for $\omega>0$ ($\omega<0$) 
[Figs. \ref{fig:U16}(c), \ref{fig:U16}(e), \ref{fig:quarterFilling}(a), and \ref{fig:quarterFilling}(c)] as in the Hubbard chain \cite{Kohno1DHub}. 
The properties of the doping-induced states and the relationship to the Mott transition are discussed in detail in Sec. \ref{sec:MottTrans}. 
\par
It should be noted that the derivation of the effective $U$, $U_{\rm eff}^\pm$ [Eq. (\ref{eq:effU})], is valid for $U$, $t_{\perp}\gg t_{\parallel}$ 
(not only $U\gg t_{\perp}\gg t_{\parallel}$) for $n\le 1/2$. 
In the case of $t_{\perp}\gg U\gg t_{\parallel}$, 
\begin{equation}
\begin{aligned}
U_{\rm eff}^+&=4t_{\perp}+&\frac{U}{2}+\frac{U^2}{16t_{\perp}},\\
U_{\rm eff}^-&=&\frac{U}{2}-\frac{U^2}{16t_{\perp}},
\end{aligned}
\end{equation}
up to $O(U^2/t_{\perp})$. 
The dimer Mott gap determined by $U_{\rm eff}^-$ at $n_{\rm eff}(=2n)=1$ is of $O(U)$ for $t_{\perp}\gg U\gg t_{\parallel}$. 
Thus, in the case of $U$, $t_{\perp}\gg t_{\parallel}$, 
the dimer Mott gap at $n=1/2$ as well as $U_{\rm eff}^-$ for $n\le 1/2$ is limited by 2$t_{\perp}$ for $U\gg t_{\perp}$ [Fig. \ref{fig:Gapn05}(a)] 
and by $U/2$ for $t_{\perp}\gg U$ [Fig. \ref{fig:Gapn05}(a) with $2t_{\perp}\leftrightarrow U/2$]. 
This implies that the effective Hubbard model with $U_{\rm eff}> 2t_{\perp}$ ($U_{\rm eff}> U/2$) is not relevant to the low-energy properties 
of a dimer Mott insulator for $U\gg t_{\perp}$ ($t_{\perp}\gg U$). 
Because $U_{\rm eff}^\pm$ is obtained in a dimer, 
the above argument would generally hold true for coupled-dimer systems \cite{KinoEffU,SeoRev} 
regardless of the lattice structure or dimensionality as long as $U$, $t_{\perp}\gg t_{\parallel}$. 
\section{Half-filling} 
\subsection{Electronic excitation} 
At half-filling ($n=1$), the ground state of the Hubbard ladder for $U\gg t_{\perp}\gg t_{\parallel}$ can be effectively approximated as 
\begin{equation}
|{\rm GS}\rangle\approx\prod_{j=1}^{L}|\psi_-\rangle_j.
\label{eq:GS}
\end{equation}
The dominant modes excited from the ground state are obtained by replacing one of the $|\psi_-\rangle$'s 
with $|B\rangle$, $|A\rangle$, $|G\rangle$, and $|F\rangle$ as 
\begin{equation}
|\overline{X(k_{\parallel})}\rangle=\frac{1}{\sqrt{L}}\sum_{j=1}^L\mathrm{e}^{i k_{\parallel}r_j}
|X\rangle_j\prod_{l\ne j}^{L}|\psi_-\rangle_l,
\label{eq:mode}
\end{equation}
where $X$ represents $B$, $A$, $G$, and $F$. 
The excitation energies up to the second order in $t_{\parallel}$ are obtained as 
\begin{equation}
\epsilon^X_{k_{\parallel}}=-2t_{\parallel}^X\cos k_{\parallel}+E_X-E_{\psi^-}
+2\xi_{X\psi^-}-2\xi_{\psi^-\psi^-},
\label{eq:1p}
\end{equation}
where $E_X$ denotes the rung energy of $|X\rangle$ (Table \ref{tbl:rung}), and 
$t_{\parallel}^B=-t_{\parallel}^F=t_{\parallel}^+$ and $t_{\parallel}^A=-t_{\parallel}^G=t_{\parallel}^-$, 
where 
\begin{equation}
t_{\parallel}^\pm=-\frac{t_{\parallel}}{2}\left(1\pm \frac{4t_{\perp}}{\sqrt{U^2+16t_{\perp}^2}}\right).
\end{equation}
Here, $\xi_{X\psi^-}$ denotes the bond energy between $|X\rangle$ and $|\psi^-\rangle$ obtained in the second-order perturbation theory, 
\begin{equation}
\begin{aligned}
\xi_{\psi^-\psi^-}=&-\frac{2t_{\parallel}^2U^2}{(U^2+16t_{\perp}^2)^{3/2}},\\
\xi_{T\psi^-}=&-\frac{4t_{\parallel}^2}{U}+\frac{2t_{\parallel}^2}{\sqrt{U^2+16t_{\perp}^2}},\\
\xi_{B\psi^-}=&-\frac{t_{\parallel}^2}{4t_{\perp}}-\frac{t_{\parallel}^2\sqrt{U^2+16t_{\perp}^2}}{8t_{\perp}U}
+\frac{\xi_{\psi^-\psi^-}}{8}\\
&+\frac{t_{\parallel}^2}{U}+\frac{t_{\parallel}^2}{\sqrt{U^2+16t_{\perp}^2}}
-\frac{2t_{\parallel}^2t_{\perp}}{U\sqrt{U^2+16t_{\perp}^2}},\\
\xi_{A\psi^-}=&\xi_{B\psi^-}|_{t_{\perp}\leftrightarrow -t_{\perp}},\\
\xi_{G\psi^-}=&\xi_{A\psi^-},\\
\xi_{F\psi^-}=&\xi_{B\psi^-},
\end{aligned}
\label{eq:bondE}
\end{equation}
where $T$ represents $T^+$, $T^-$, or $T^0$. 
\par
We can also construct two-particle states by using $X(=B$, $A$, $G$, and $F)$ and 
$Y(=T^+$, $T^-$, or $T^0)$ as 
\begin{eqnarray}
|\overline{XT(k_{\parallel};p_{\parallel})}\rangle&=&\frac{1}{\sqrt{L(L-1)}}\sum_{m\ne n}
\mathrm{e}^{i(k_{\parallel}-p_{\parallel})r_m}\mathrm{e}^{i p_{\parallel}r_n}\nonumber\\
&&|X\rangle_m|Y\rangle_n\prod_{l\ne m,n}^{L}|\psi_-\rangle_l, 
\label{eq:cont}
\end{eqnarray}
whose effective excitation energies for $L\rightarrow\infty$ are obtained as 
\begin{eqnarray}
\epsilon^{XT}_{k_{\parallel};p_{\parallel}}&=&-2t_{\parallel}^X\cos(k_{\parallel}-p_{\parallel})
+J^{\rm eff}_{\parallel}\cos p_{\parallel}+E_X-2E_-\nonumber\\
&&+2\mu+2\xi_{X\psi^-}+2\xi_{T\psi^-}-4\xi_{\psi^-\psi^-},
\label{eq:2p}
\end{eqnarray}
where 
\begin{equation}
J^{\rm eff}_{\parallel}=\frac{8t_{\parallel}^2}{U}-\frac{4t_{\parallel}^2}{\sqrt{U^2+16t_{\perp}^2}}, 
\end{equation}
which reduces to $J^{\rm eff}_{\parallel}\approx J_{\parallel}$ for $U\gg t_{\perp}$. 
\par
\begin{figure}
\includegraphics[width=8.3cm]{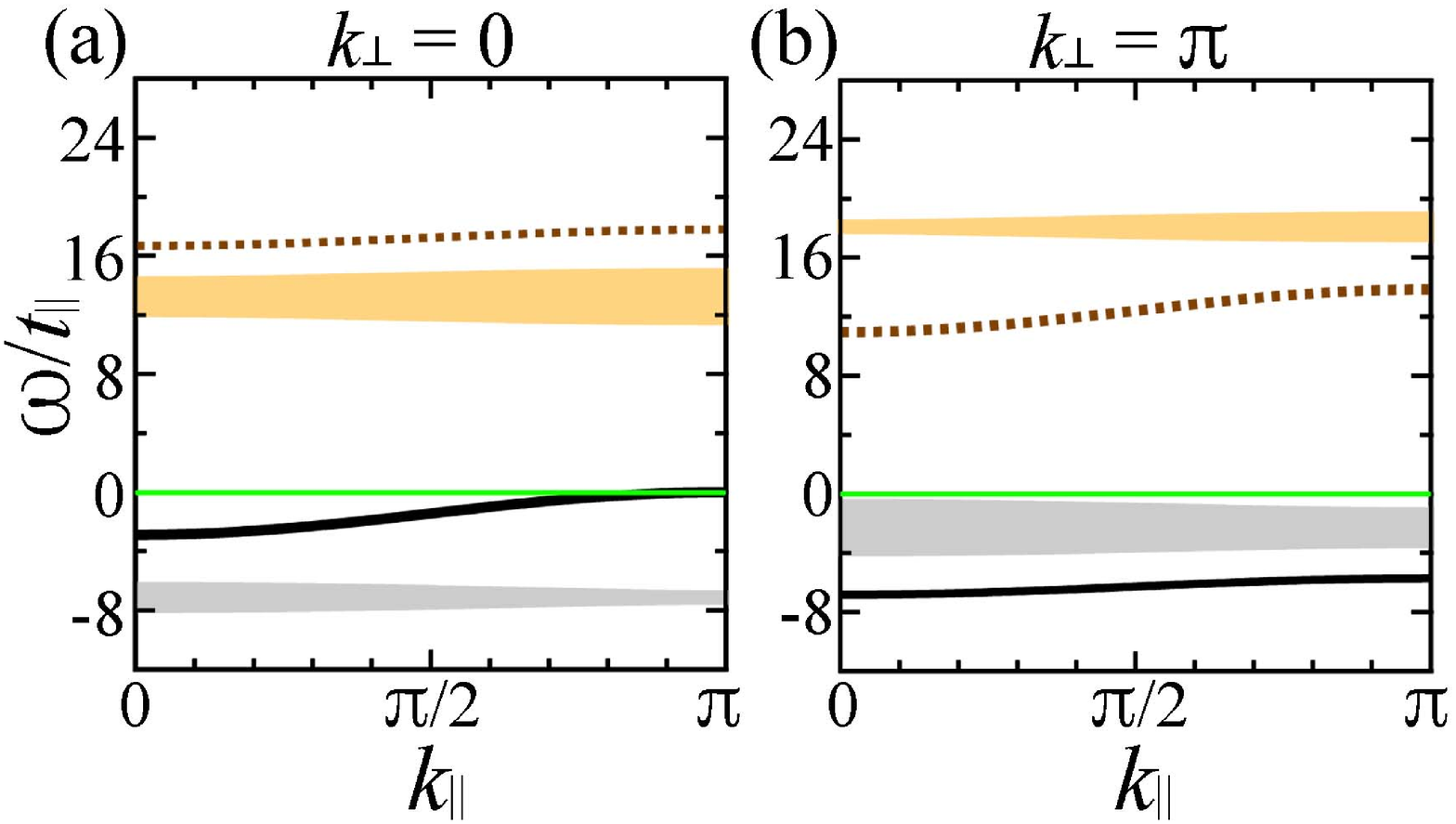}
\caption{Dispersion relation of electronic excitations at $n=1$ for $U/t_{\parallel}=16$ and $t_{\perp}/t_{\parallel}=2$ at $k_{\perp}=0$ [(a)] and $\pi$ [(b)]. 
In (a), $\omega=\epsilon^G_{k_{\parallel}}$ (dotted brown curve), $-\epsilon^B_{k_{\parallel}}$ (solid black curve), 
$\epsilon^{FT}_{k_{\parallel};p_{\parallel}}$ (light orange region), and $-\epsilon^{AT}_{k_{\parallel};p_{\parallel}}$ (light gray region). 
In (b), $\omega=\epsilon^F_{k_{\parallel}}$ (dotted brown curve), $-\epsilon^A_{k_{\parallel}}$ (solid black curve), 
$\epsilon^{GT}_{k_{\parallel};p_{\parallel}}$ (light orange region), and $-\epsilon^{BT}_{k_{\parallel};p_{\parallel}}$ (light gray region). 
The green lines indicate $\omega=0$. The chemical potential $\mu$ is set so that $\epsilon^B_{\pi}=0$.} 
\label{fig:HF}
\end{figure}
The dominant modes and the continua carrying considerable spectral weights in Figs. \ref{fig:U16}(h) and \ref{fig:U16}(p) 
can basically be identified with the above modes [Eqs. (\ref{eq:mode}) and (\ref{eq:1p}); dotted brown curves and solid black curves in Fig. \ref{fig:HF}] 
and the two-particle states [Eqs. (\ref{eq:cont}) and (\ref{eq:2p}); light orange regions and light gray regions in Fig. \ref{fig:HF}]. 
\subsection{Spin excitation} 
\begin{figure}
\includegraphics[width=8.3cm]{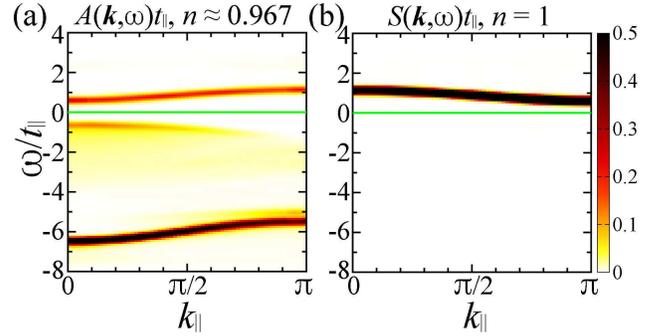}
\caption{(a) $A({\bm k},\omega)t_{\parallel}$ for $k_{\perp}=\pi$ at $n\approx 0.967$ 
for $U/t_{\parallel}=16$ and $t_{\perp}/t_{\parallel}=2$ obtained using the non-Abelian DDMRG method 
[closeup of Fig. \ref{fig:U16}(o) near the Fermi level]. 
(b) $S({\bm k},\omega)t_{\parallel}$ for $k_{\perp}=\pi$ at $n=1$ 
for $U/t_{\parallel}=16$ and $t_{\perp}/t_{\parallel}=2$ obtained using the non-Abelian DDMRG method. 
The green lines indicate $\omega=0$. Gaussian broadening with a standard deviation of $0.1t_{\parallel}$ has been used.} 
\label{fig:SkwAkw}
\end{figure}

The spin excited state is similarly obtained as in Eq. (\ref{eq:mode}) with $X=T^+$, $T^-$, or $T^0$, 
whose excitation energy up to the second order in $t_{\parallel}$ can be obtained as
\begin{equation}
\epsilon^{\rm spin}_{k_{\parallel}}=J_{\parallel}^{\rm eff}\cos k_{\parallel}-E_-
+2\xi_{T\psi^-}-2\xi_{\psi^-\psi^-}.
\label{eq:spin}
\end{equation}
This excitation well explains the mode in $S({\bm k},\omega)$ for $k_{\perp}=\pi$ at half-filling [Fig. \ref{fig:SkwAkw}(b)]. 
\section{Mott transition} 
\label{sec:MottTrans}
\subsection{Doping-induced states} 
\label{sec:DIS}
The most remarkable spectral feature is the emergence of electronic states in the Mott gap by doping a Mott insulator [Figs. \ref{fig:U16}(o) and \ref{fig:U16}(p)]. 
From the low-electron-density side, the dominant mode for $k_{\perp}=\pi$ in the low-electron-density regime loses spectral weight and 
disappears at half-filling [Figs. \ref{fig:U16}(i)--\ref{fig:U16}(p)], 
but its dispersion relation remains dispersing until the Mott transition occurs [Figs. \ref{fig:U16}(o) and \ref{fig:SkwAkw}(a)]. 
In the small-doping limit, the dispersion relation reduces to the magnetic dispersion relation 
shifted by the Fermi momentum ${\bm k}_{\rm F}=(\pi,0)$ (Fig. \ref{fig:SkwAkw}). 
Thus, the antibonding band in the low-electron-density regime gradually loses spectral weight with the electron density 
and eventually leads to the magnetic excitation at half-filling. 
This implies that the charge degrees of freedom freeze, 
whereas the spin degrees of freedom remain active toward the Mott transition \cite{Kohno1DHub,Kohno2DHub,KohnoDIS,KohnoRev}. 
\par
From the Mott-insulator side, this feature can be described as follows: 
the spin excited states at half-filling appear in the electron-addition spectrum with the dispersion relation 
shifted by the Fermi momentum upon doping a Mott insulator 
because the charge characteristic is added to the spin excited states by doping \cite{Kohno1DHub,Kohno2DHub,KohnoDIS,KohnoRev}. 
A simple explanation for this feature has been given using effective eigenstates of the $t$-$J$ ladder 
as well as based on general arguments on quantum numbers in Ref. \cite{KohnoDIS}. 
Here, we briefly review the explanation in the case of the Hubbard ladder. 
The ground state at half-filling for $U\gg t_{\perp}\gg t_{\parallel}$ can be effectively approximated as in Eq. (\ref{eq:GS}). 
The spin excited states are obtained as $|\overline{T(k_{\parallel})}\rangle$ ($T=T^+, T^-$, or $T^0$) [Eq. (\ref{eq:mode})] 
whose dispersion relation is expressed as $\omega=\epsilon^{\rm spin}_{k_{\parallel}}$ [Eq. (\ref{eq:spin}); Fig. \ref{fig:SkwAkw}(b)]. 
The one-hole-doped ground state is essentially $|\overline{B(\pi)}\rangle$, which has momentum ${\bm k}=(\pi,0)$. 
In the small-doping limit, the chemical potential is set so that the energy of the one-hole-doped ground state is zero. 
When an electron with momentum $(p_{\parallel},\pi)$ is added to the one-hole-doped ground state, 
the obtained state has overlap with the spin excited state at half-filling $|\overline{T(p_{\parallel}+\pi)}\rangle$. 
Thus, $A((p_{\parallel},\pi),\omega)$ exhibits a mode along $\omega=\epsilon_{p_{\parallel}+\pi}^{\rm spin}$ [Eq. (\ref{eq:spin}); Fig. \ref{fig:SkwAkw}(a)]. 
This argument shows that the spin excited states at half-filling appear in the electron-addition spectrum 
with the dispersion relation shifted by the Fermi momentum upon doping a Mott insulator. 
\par
This feature of the doping-induced states has also been pointed out in the Hubbard chain \cite{Kohno1DHub}, two-dimensional (2D) Hubbard model \cite{Kohno2DHub}, 
$t$-$J$ chain \cite{Kohno1DtJ}, 2D $t$-$J$ model \cite{Kohno2DtJ}, and $t$-$J$ ladder \cite{KohnoDIS}, 
as well as in a system with antiferromagnetic order \cite{KohnoAF}. 
In the Hubbard ladder considered in this paper, the mode of the doping-induced sates has an energy gap 
because the spin excitation at half-filling has an energy gap (Fig. \ref{fig:SkwAkw}). In the case where the spin excitation is gapless in a Mott insulator, 
the mode of the doping-induced states should be gapless, as shown in Fig. \ref{fig:quarterFilling}, and 
in the Hubbard and $t$-$J$ chains, and the 2D Hubbard and $t$-$J$ models \cite{Kohno1DHub,Kohno2DHub,Kohno1DtJ,Kohno2DtJ,KohnoDIS,KohnoAF,KohnoRev,KohnoSpin}. 
\par
The emergence of electronic states upon doping a Mott insulator was recognized soon after the discovery of cuprate high-temperature superconductors \cite{DagottoRMP,Eskes,DagottoDOS}. 
However, interpretations are controversial. 
Various interpretations other than the above interpretation have been proposed 
primarily for the 2D Hubbard model \cite{SakaiImadaPRL,ImadaCofermionPRL,PhillipsMottness,EderOhta2DHub,EderOhtaIPES}, 
suggesting that the mode of the doping-induced states is essentially separated by a (pseudo-)gap from the low-energy band 
even though the Mott insulator exhibits gapless spin excitation \cite{SakaiImadaPRL,ImadaCofermionPRL,PhillipsMottness,EderOhta2DHub,EderOhtaIPES}. 
In contrast, the interpretation described above as well as in Refs. \cite{Kohno1DHub,Kohno2DHub,Kohno1DtJ,Kohno2DtJ,KohnoDIS,KohnoAF,KohnoRev,KohnoSpin} 
can naturally and collectively explain the behavior of the doping-induced states in the Hubbard and $t$-$J$ chains \cite{Kohno1DHub,Kohno1DtJ} 
and the Hubbard and $t$-$J$ ladders \cite{KohnoDIS} (Figs. \ref{fig:quarterFilling} and \ref{fig:SkwAkw}) as well as in the 2D Hubbard and $t$-$J$ models \cite{Kohno2DHub,Kohno2DtJ,KohnoRev}, 
which implies that this interpretation captures the essence of the Mott transition. 
\subsection{What characterizes the Mott transition} 
To discuss how to characterize the Mott transition, the definition of the Mott transition must be clarified. 
In particular, the definition should be what can distinguish the Mott transition from the transition between a metal and a band insulator. 
Hence, a clear distinction between a Mott insulator and a band insulator is required. 
\par
One might consider that a Mott insulator could be defined by the value of the charge gap; 
if the charge gap is primarily determined by the Coulomb repulsion [$O(U)$ in the Hubbard model], the insulating state could be regarded as a Mott insulator. 
However, as shown in Sec. \ref{sec:quarterFilling}, even though the value of the charge gap is limited by the intradimer hopping for $U\gg t_{\perp}\gg t_{\parallel}$, 
the insulating state at quarter-filling should be regarded as a Mott insulator 
because it is essentially the same as the Mott insulator of the Hubbard chain (Fig. \ref{fig:quarterFilling}) [Sec. \ref{sec:quarterFilling}]. 
Thus, a Mott insulator is not necessarily well-defined in terms of the value of the charge gap. 
\par
Another definition could be that a Mott insulator is an insulator exhibiting gapless spin excitation. 
Such an insulator is not a band insulator because the spin gap is basically equal to the charge gap in a band insulator. 
However, this definition appears too narrow; in practice, a system having low-energy spin excitation with a large charge gap 
is generally called a Mott insulator regardless of whether a small spin gap opens or not. 
Hence, a Mott insulator can be better defined as an insulator with $\Delta_{\rm s}\ll\Delta_{\rm c}$ (or $\Delta_{\rm s}<\Delta_{\rm c}$ if necessary), 
where $\Delta_{\rm s}$ and $\Delta_{\rm c}$ denote 
the lowest excitation energies for spin and charge, respectively. 
This implies that a Mott insulator can be defined in terms of the spin-charge separation ($\Delta_{\rm s}\ne \Delta_{\rm c}$) \cite{KohnoDIS}. 
In 1D systems, the spin-charge separation is considered to occur even in a metallic phase: 
The properties in the low-energy limit are described in terms of spin and charge excitations independent of each other 
with different velocities rather than electronlike quasiparticles \cite{HaldaneTLL,TomonagaTLL,LuttingerTLL,MattisLiebTLL}. 
In a Mott insulator, regardless of the lattice structure or dimensionality, 
the spin-charge separation ($\Delta_{\rm s}\ll \Delta_{\rm c}$) occurs more clearly than in a 1D metal [$\Delta_{\rm s}\ne \Delta_{\rm c}=O(1/L)$]. 
\par
If a Mott insulator is characterized by $\Delta_{\rm s}\ll\Delta_{\rm c}$, what characterizes the Mott transition should also reflect it. 
If only the ground-state properties are considered, the Mott transition in a dimerized system, 
such as the Hubbard ladder for $U\gg t_{\perp}\gg t_{\parallel}$ at half-filling, 
is essentially the same as the transition to a band insulator [Figs. \ref{fig:U16}(g) and \ref{fig:U16}(h)]. 
However, reflecting $\Delta_{\rm s}\ll\Delta_{\rm c}$ of the Mott insulator [Figs. \ref{fig:U16}(h), \ref{fig:U16}(p), and \ref{fig:SkwAkw}(b)], 
electronic states exhibiting the momentum-shifted magnetic dispersion relation emerge in the Mott gap by doping [Figs. \ref{fig:U16}(o) and \ref{fig:SkwAkw}(a)]. 
This characteristic of the Mott transition reflects the characteristic of the Mott insulator (spin-charge separation: $\Delta_{\rm s}\ll\Delta_{\rm c}$) and does not appear in the transition from a band insulator. 
Hence, this characteristic should be general and fundamental to the Mott transition \cite{Kohno1DHub,Kohno2DHub,Kohno1DtJ,Kohno2DtJ,KohnoDIS,KohnoAF,KohnoRev,KohnoSpin}. 
The above argument implies that the Mott transition is characterized by the doping-induced states that exhibit the momentum-shifted magnetic dispersion relation 
rather than critical exponents or order parameters. 
\section{Summary} 
The emergence, disappearance, and spectral-weight transfer of electronic states are illustrated 
in the Hubbard ladder in the strong repulsion and strong intrarung hopping regime. 
The dominant modes in the low-electron-density regime significantly lose spectral weight as the electron density increases to half-filling, 
whereas the emergent modes in the low-electron-density regime become dominant at half-filling; 
the dominant modes in the low-electron-density regime and those at half-filling are different in origin. 
\par
One of the emergent modes, which has an energy of the order of the intradimer hopping, significantly gains spectral weight and governs the dimer Mott physics at quarter-filling; 
the dimer Mott gap is limited by the intradimer hopping even in the strong repulsion regime. 
\par
In contrast, one of the dominant modes in the low-electron-density regime, which originates from a noninteracting band, 
gradually loses spectral weight as the electron density increases 
and completely disappears at half-filling. 
However, the dispersion relation remains dispersing until the Mott transition occurs; 
the dispersion relation reduces to the magnetic dispersion relation shifted by the Fermi momentum in the small-doping limit. 
Thus, the mode originating from a noninteracting band continuously leads to the mode of the spin excitation at half-filling. 
\par
These features would be basically true for general coupled-dimer systems regardless of the lattice structure or dimensionality as long as $U\gg t_{\perp}\gg t_{\parallel}$. 
As the dimer Mott gap is limited by the intradimer hopping, 
a Mott insulator is not necessarily well-defined as an insulator having a charge gap of the order of the Coulomb repulsion; 
a Mott insulator is better characterized in terms of the existence of spin excitation whose energy is much lower than the charge gap, i.e., the spin-charge separation. 
By reflecting this characteristic of a Mott insulator, the Mott transition can be characterized: 
The spin excited states in a Mott insulator emerge in the Mott gap as electronic excitation by doping the Mott insulator, 
exhibiting the momentum-shifted magnetic dispersion relation. 
This feature does not appear in the transition from a band insulator and should be general and fundamental to the Mott transition. 
\par
In the small-$t_{\perp}/t_{\parallel}$ regime, the spectral-weight distribution can be affected by band hybridization. 
Nevertheless, the overall spectral features such as emergence and disappearance of spectral weight 
should generally appear in strongly correlated systems. 
\par
The emergence, disappearance, and spectral-weight transfer, which have almost been overlooked in band theory and Fermi-liquid theory, 
play particularly important roles in understanding the physics around the Mott transition, such as the origin of the Mott gap and the doping-induced states. 
This would be one of the reasons why the electronic properties near the Mott transition have appeared elusive from the conventional viewpoints. 
This paper also brings an unconventional perspective to electronic bands. 
In a conventional band picture, electronic bands are usually identified as those primarily originating from atomic orbitals; 
the number of bands is considered invariant regardless of the electron density as long as symmetry breaking does not occur. 
However, as shown in this paper, a noninteracting band at zero electron density, which corresponds to a conventional band, can disappear at the Mott transition, 
whereas emergent electronic bands in the low-electron-density regime can become dominant near the Mott transition. 
The number of electronic bands carrying significant spectral weight can vary depending on the electron density even without symmetry breaking in strongly correlated systems. 
Experimental confirmation of these features over a wide energy and electron-density regime 
as well as applications of the emergence of electronic states to electronic or optical devices is desired. 
\begin{acknowledgments}
The author would like to thank S. Uji and S. Tsuda for helpful discussions. 
This work was supported by JSPS KAKENHI Grant No. JP26400372 and the JST-Mirai Program Grant No. JPMJMI18A3, Japan. 
The numerical calculations were partly performed on the supercomputer at the National Institute for Materials Science. 
\end{acknowledgments}

\end{document}